\documentclass[prd,aps,nofootinbib,11pt,preprint]{revtex4}

\newcommand{\checked}[1]{}

\usepackage{graphicx}
\usepackage{amsmath}
\usepackage[colorlinks=true,linktocpage=true,linkcolor=blue,citecolor=blue]{hyperref}

\newcommand{\beq}{\begin{equation}}
\newcommand{\eeq}{\end{equation}}
\newcommand{\bqa}{\begin{eqnarray}}
\newcommand{\eqa}{\end{eqnarray}}
\def\simge{\mathrel{
    \rlap{\raise 0.511ex \hbox{$>$}}{\lower 0.511ex \hbox{$\sim$}}}}
\def\simle{\mathrel{
    \rlap{\raise 0.511ex \hbox{$<$}}{\lower 0.511ex \hbox{$\sim$}}}}

\begin{document}

\title{\bf Bulk viscous corrections to screening and damping in QCD at high temperatures}

\author{Qianqian Du$^{a}$, Adrian Dumitru$^{b,c}$, Yun Guo$^{a}$, and Michael Strickland$^{d}$}
\affiliation{
$^a$Department of Physics, Guangxi Normal University,
Guilin, 541004, China\\
$^b$Department of Natural Sciences, Baruch College, CUNY,
17 Lexington Avenue, New York, NY 10010, USA\\
$^c$The Graduate School and University Center, The City
  University of New York,\\
  365 Fifth Avenue, New York, NY 10016, USA\\
$^d$Department of Physics, Kent State University, 206B Smith Hall, Kent, OH 44240, USA
}

\begin{abstract}
Non-equilibrium corrections to the distribution functions of quarks
and gluons in a hot and dense QCD medium modify the ``hard thermal
loops'' (HTL). The HTLs determine the retarded, advanced, and symmetric
(time-ordered) propagators for gluons with soft momenta as well as the
Debye screening and Landau damping mass scales. We compute such
corrections to a thermal as well as to a non-thermal fixed point. The
screening and damping mass scales are sensitive to the bulk pressure
and hence to (pseudo-) critical dynamical scaling of the bulk
viscosity in the vicinity of a second-order critical point. This could be
reflected in the properties of quarkonium bound states in the deconfined
phase and in the dynamics of soft gluon fields.
\end{abstract}

\maketitle

%-----------------------------------------------------------------------
\section{Introduction} \label{sec:Intro}

In order to understand the physics of the quark-gluon plasma (QGP)
generated in ultrarelativistic heavy-ion collisions a first step is to
understand the dynamics of the high temperature phase of QCD.
At extremely high temperatures $T \gg \Lambda_{\rm QCD}$ the system
can be described as a weakly-interacting gas of quark and gluon
quasiparticles which can be understood systematically using
hard-thermal loop (HTL) resummation
\cite{Weldon:1982aq,Braaten:1989mz,Frenkel:1989br,Braaten:1991gm}.
Such a picture has been quite successful in describing the
thermodynamics of the QGP down to temperatures on the order of $T \sim
250$ MeV, particularly when considering the various quark
susceptibilities \cite{Haque:2014rua,Andersen:2011sf,Andersen:1999fw}.
The resulting picture is one in which the QGP is comprised of
quasiparticle-like excitations which experience Landau-damping in a
similar manner as electromagnetic plasma waves.  In the
high-temperature equilibrium limit there is only one scale, the Debye
mass $m_D \sim gT$, which enters the quark and gluon self energies.
For both quarks and gluons, in equilibrium the retarded two-point
function has two real time-like ($\omega>k$) poles corresponding to
propagating soft modes (plasmon/plasmino and transverse/longitudinal
for quarks and gluons, respectively) and a space-like ($\omega < k$)
cut which results in Landau-damping of soft quark and gluon modes.
Such modifications of the two-point functions are reflected in
analogous hard-thermal-loop modifications to all soft $n$-point
functions which must be taken into account in order to maintain the
explicit gauge-invariance of the soft resummation program
\cite{Braaten:1991gm}.

Knowledge of the HTL-resummed gluon self-energy allows one to compute
quantities such as the inter-quark potential in the heavy quark limit
\cite{Laine:2006ns}. The resulting heavy-quark potential is
complex-valued, with the real part of the potential taking the form of
a Debye-screened Coulomb potential which reflects color-screening in
the quark-gluon plasma (QGP) and the imaginary part being related to
the in-medium decay width of heavy quark bound states.  As interesting
as this is in and of itself, if one is interested in QGP
phenomenology, then one must incorporate non-equilibrium corrections
to the heavy-quark potential. This requires input information about
the analogously resummed non-equilibrium quark and gluon
self-energies.  In the early stages of a heavy-ion collision when the
temperature is highest and the expansion is highly anisotropic the
most important non-equilibrium correction in the QGP stems from finite
shear viscosity of the plasma, $\eta$.

When $\eta$ is non-zero, the rapid longitudinal expansion of the QGP
created in relativistic heavy ion collisions results in
anisotropies in the diagonal components of the energy-momentum tensor
in the local rest frame and, in a kinetic-theory framework, this
translates into momentum-space anisotropies in the quark and gluon
distribution functions~\cite{Strickland:2013uga,Strickland:2014pga}.
As a result, one must revisit the calculation of the heavy-quark
potential, taking into account these momentum-space
anisotropies~\cite{Dumitru:2007hy,Dumitru:2009fy,Burnier:2009yu}.
Such calculations have led to detailed phenomenological calculations
of the expected level of heavy quarkonium suppression generated in
high-energy heavy-ion collisions
\cite{Strickland:2011mw,Strickland:2011aa,Krouppa:2015yoa,Krouppa:2016jcl}.
These papers have demonstrated that it is necessary to include the
shear correction to the heavy-quark potential when computing
heavy-quark suppression.

In recent years, attention has broadened to include and fit other
transport coefficients in the QGP with the most obvious candidate
being the bulk viscosity, $\zeta$. It has been demonstrated that
self-consistent inclusion of bulk viscous effects improves the
agreement of hydrodynamical model predictions with experimental data,
see e.g. \cite{Ryu:2015vwa}.  The bulk viscosity $\zeta(T)$ in QCD at
very high temperatures $T\gg\Lambda_{\rm QCD}$ has been computed to
leading order in the coupling in Ref.~\cite{Arnold:2006fz}. They find
that it is very small indeed since $\zeta$ is proportional to the {\em
  square} of the deviation from conformality given by the
$\beta$-function. This leads to $\zeta/\eta\sim \alpha_s^4$
(neglecting logarithms of the inverse coupling).

On the other hand, it is known from the lattice that the trace anomaly
of QCD, expressed as energy density minus three times the pressure,
grows large at $T\sim\Lambda_{\rm QCD}$~\cite{Bazavov:2014pvz}.  Thus,
it has been suggested in the literature that the bulk viscosity to
entropy density ratio should increase, too, as the temperature
approaches the confinement-deconfinement
temperature~\cite{KharzeevTuchin}.  In this paper we analyze the high-temperature
weakly-coupled phase of the QGP and try to assess the impact of bulk-viscous corrections
on the heavy-quark potential.  While our weak-coupling analysis
may not apply for $T\simeq \Lambda_{\rm QCD}$, nevertheless it is
clearly of interest to obtain a baseline expectation for bulk-viscous
effects on screening and damping from (resummed) weakly coupled QCD.

Bulk viscous corrections are expected to grow large also in the
vicinity of a second order critical point; this could be realized in
hot QCD either by tuning of the quark masses~\cite{qmass} or perhaps
by introducing a baryon charge asymmetry~\cite{muB_critical_point}.
Due to critical slowing down the bulk viscosity should
diverge~\cite{Moore:2008ws} $\zeta\sim \xi^z$ where $\xi\to\infty$ is
the correlation length and $z$ is a dynamical critical
exponent. However, since the relaxation time in the critical region of
the bulk pressure diverges as well, in heavy-ion collisions its
magnitude relative to the ideal pressure should not be much greater
than $\sim 1$~\cite{Monnai:2016kud}.

Additionally, we mention that there is shear-bulk coupling in
non-conformal relativistic viscous hydrodynamics derived from kinetic
theory via the 14 moment approximation~\cite{Denicol:2014vaa}. Due to
this coupling, a large shear pressure may induce significant
bulk-viscous corrections, and possibly even invert their
sign~\cite{BulkShearKinetic}.

To compute the gluon self energy in the hard loop approximation we
require the phase space distributions of the particles in the medium.
In the local rest frame, we take them to be
\beq
f({\bf p}) = f_{\rm id}(p) + \delta_{\rm bulk} f(p) + \delta_{\rm shear}
f({\bf p})~.
\label{eq:dis}
\eeq
Here, $f_{\rm id}(p)$ is an isotropic reference distribution when
non-equilibrium corrections are absent. This would normally correspond
to thermal Fermi-Dirac or Bose-Einstein distributions, respectively,
if the ``ideal'' reference is the thermal fixed point; see
Section~\ref{sec:thermalfp}. In Sec.~\ref{sec:nonthermalfp} on the
other hand we shall choose a non-thermal fixed point parameterized by
a mass-like (scalar field) distortion of the ideal gas distributions,
with $m\sim T$.

The corrections $\delta f$ in Eq.~(\ref{eq:dis}) correspond to non-equilibrium
corrections. We denote the isotropic correction $\delta_{\rm bulk}
f(p)$ as a bulk-viscous correction while the anisotropic part
$\delta_{\rm shear} f({\bf p})$ is analogous to shear. However, we do
not assume that these corrections are parametrically suppressed. The
corrections to the real and imaginary parts of the HTL resummed gluon
propagator due to $\delta_{\rm shear} f({\bf p})$ have been worked out
in Refs.~\cite{Dumitru:2007hy,Dumitru:2009fy}.  Here, we focus on bulk
viscous corrections instead. Unlike the thermal distribution functions
non-equilibrium corrections are not universal and so we work out
explicit expressions for two different examples in
Sections~\ref{sec:thermalfp} and~\ref{sec:nonthermalfp}, respectively.

Throughout the manuscript we use natural units $\hbar=c=k_B=1$ and a
``mostly minus'' ($+---$) metric. Capital letters denote four-momenta
while lower caps letters are three-momenta.

%-----------------------------------------------------------------------
\section{Bulk-viscous corrections about a thermal fixed
  point} \label{sec:thermalfp}
In this section we compute the temporal component of the gluon self energies for massless
thermal particles. The retarded gluon self energy in the real time
formalism is given by~\cite{Mrowczynski:2000ed}\footnote{Formally, this expression can be obtained from the corresponding result in Ref.~\cite{Carrington:1997sq} at zero chemical potential by the simple
replacement $f({\bf k})\rightarrow(f^+({\bf k})+f^-({\bf k}))/2$. However, this is in general not true for the symmetric self energy. See Appendix.}
\begin{eqnarray}
 \Pi_R(P)=\frac{2 \pi N_f g^2}{(2\pi)^4}\int k d k d \Omega_k\, (f^+_F({\bf
k})+f^-_F({\bf k})) \frac{1-({\hat{\bf k}}\cdot {\hat{\bf p}})^2}{({\hat{\bf
k}}\cdot {\hat{\bf p}}+\frac{p_0+i\, \epsilon}{p})^2}~.
\label{general:re}
\end{eqnarray}
This expression accounts for the contribution due to $N_f$ (massless)
quark loops. Three-momenta with a hat denote unit vectors. The distribution
function may have any non-equilibrium form so long as the dominant
contribution is from hard loop momenta of order $T$ so that the HTL
approximation is applicable. In the thermal equilibrium case, the distribution function for (anti-)quarks with chemical potential $\mu$ is given by
\begin{eqnarray}
n^{\pm}_F(k)=\frac{1}{{\rm {exp}}[(k\mp \mu)/T]+1}~.
\label{general:dis}
\end{eqnarray}

In the absence of non-equilibrium corrections the ideal distribution $f_{{\rm {id}}}(k)$
function is given by $n^{\pm}_F(k)$ and we have
\begin{equation}
 \Pi_{R}^{\rm {id}}(P)=N_f \frac{g^2
   T^2}{6}
 \left(1+\frac{3 \tilde{\mu}^2}{\pi^2}\right)
 \left (\frac{p_0}{2 p}\ln \frac{p_0+p+i\epsilon}
{p_0-p+i\epsilon} -1\right)~,\label{eq:Pi00_q_ideal}
\end{equation}
where the dimensionless quantity $\tilde{\mu}$ is defined as $\tilde{\mu}\equiv \frac{\mu}{T}$. The contribution due to a gluon loop has the same structure as
Eq.~(\ref{general:re}) but with $f_{\rm id}(k)$ now a Bose distribution.
In equilibrium,
\begin{equation}
 \Pi_{R}^{\rm {id}}(P)=2N_c \frac{g^2
T^2}{6}\left (\frac{p_0}{2 p}\ln \frac{p_0+p+i\epsilon}
{p_0-p+i\epsilon} -1\right)~.\label{eq:Pi00_g_ideal}
\end{equation}

The symmetric (time ordered) self energy due to $N_f$ quark loops is given
by\footnote{See Appendix for details.}
\begin{eqnarray}
 \Pi _F(P)= 4 i N_f g^2 \pi^2\int
\frac{k^2dk}{(2 \pi )^3}\sum_{i=\pm}f^i_F(k)(f^i_F(k)-1)\frac{2}{p}\Theta(p^2-p_0^2)\, .
\label{general:sy}
\end{eqnarray}
In equilibrium,
\beq
 \Pi _{F}^{\rm {id}}(P) = -2 \pi i\,N_f \frac{g^2
T^2}{6} \left(1+\frac{3 \tilde{\mu}^2}{\pi^2}\right)
\frac{T}{p}\Theta(p^2-p_0^2)~.  \label{eq:Pi_F_eq_quark}
\eeq
For a thermal gluon loop one replaces $f^i_F(k)(1-f^i_F(k))$ in Eq.~(\ref{general:sy}) by $f_B(k)(1+f_B(k))$ which leads to
\begin{equation}
 \Pi _{F}^{\rm {id}}(P)=-2 \pi i\,2 N_c \frac{g^2
T^2}{6} \frac{T}{p}\Theta(p^2-p_0^2)\, .  \label{eq:Pi_F_eq_glue}
\end{equation}
From the above results we see that at the thermal fixed point the
modification of the mass scales due to the quark-chemical potential
$\mu$ is exactly the same for both retarded (advanced) and
symmetric gluon self energies.

%-----------------------------------------------------------------
%\section{Bulk viscous correction to gluon self energies}

We now determine the non-equilibrium corrections to the expressions above.
We assume that the bulk viscous correction to the local thermal distribution
function takes the form
\beq
\delta_{\rm bulk}    f(k)=
\left(\frac{k}{T}\right)^a \Phi\, f_{\rm id}(k)(1\pm
f_{\rm id}(k))~. \label{bulk:discorrection}
\eeq
Here, the ``$+$" sign is for a Bose distribution while the ``$-$" sign
applies in case of a Fermi distribution.  $\Phi$ is proportional to
the bulk pressure (divided by the ideal pressure) and $a$ is a
constant. We require that $a>0$ to ensure that the dominant
contribution to the retarded self energy is from hard (gluon) loop
momenta, $k\sim T$. To see this note that in the massless limit the
Bose distribution for $k\ll T$ behaves as $f_B(k)\sim T/k$ and so
$f_B(k) (1+f_B(k)) \sim (T/k)^2$. The ``hard gluon loop'' from
Eqs.~(\ref{general:re},\ref{eq:PI_Rgen}), with $f(k)$ replaced by
$\delta_{\rm bulk}f(k)$, is insensitive to soft momenta $k\ll T$ if
$a>0$.  The bulk viscous correction to the symmetric self energy at
${\cal{O}} (\Phi^2)$ involves the fourth power of the distribution
function and so we have to impose a more stringent bound, $a>1/2$, in
order to employ the HTL approximation, see below. We note that these
bounds on $a$ correspond to the regime of applicability of HTL power
counting but may in principle be violated in certain non-equilibrium
scenarios.

We shall also assume that $|\Phi|\gg g^2$ so that two-loop corrections
to the gluon self energy are negligible. In fact,
since~(\ref{bulk:discorrection}) is an ad-hoc schematic model for the
non-equilibrium correction we may assume that it applies even at
$|\Phi|\sim 1$.

Since $f_{\rm id}(p) + \delta_{\rm bulk} f(p)$ is isotropic
Eq.~(\ref{general:re}) simplifies to
\beq
\Pi_R(P)=\frac{2 \pi N_f g^2}{(2\pi)^4}\int k d k \,
(f^+_F(k)+f^-_F(k)) \;
\int d \Omega_k \frac{1-({\hat{\bf k}}\cdot {\hat{\bf p}})^2}{({\hat{\bf
k}}\cdot {\hat{\bf p}}+\frac{p_0+i\, \epsilon}{p})^2}~.
\label{eq:PI_Rgen}
\eeq
Hence, the dependence on the frequency $p_0$ and on the momentum $p$
is the same as in equilibrium,
c.f.\ Eqs.~(\ref{eq:Pi00_q_ideal},\ref{eq:Pi00_g_ideal}). Specifically,
for our distribution function~(\ref{bulk:discorrection}) this
expression gives
\beq
\delta_{\rm {bulk}}    \Pi_R(P)=
c_R^{(q)}(a,\tilde{\mu}) \, \Phi N_f \frac{g^2
T^2}{6}\left(1+\frac{3 \tilde{\mu}^2}{\pi^2}\right)
\left (\frac{p_0}{2 p}\ln \frac{p_0+p+i\epsilon}
{p_0-p+i\epsilon} -1\right)~.   \label{eq:Pi00_bulk_quark}
\eeq
A similar correction is obtained for the contribution due to a gluon
loop,
\beq
\delta_{\rm{bulk}}    \Pi_R(P)=c_R^{(g)}(a) \, \Phi \,
2N_c \frac{g^2
T^2}{6}\left (\frac{p_0}{2 p}\ln \frac{p_0+p+i\epsilon}
{p_0-p+i\epsilon} -1\right)~.   \label{eq:Pi00_bulk_glue}
\eeq
The dimensionless numbers $c_R^{(q)}(a,\tilde{\mu})$ and
$c_R^{(g)}(a)$ are given by
\beq
c^{(q,g)}_R(a,\tilde{\mu}) = \frac{1}{\Phi}\frac{\int k d k \, \delta_{\rm bulk}f(k)}
{\int k d k \, f_{\rm id}(k)} =
\begin{cases}
  \frac{- 6\Gamma(2+a)[{\rm Li}_{(1+a)}(- e^{-\tilde{\mu}})+{\rm Li}_{(1+a)}(- e^{\tilde{\mu}})]}{\pi^2+3 \tilde{\mu}^2} & \text{(fermion)}~, \\
  \frac{6}{\pi^2}\Gamma(2+a)\zeta(1+a) & \text{(boson)}~,
\end{cases}
\label{eq:cRqwu}
\eeq
where  ${\rm Li}_n(z)$ denotes the polylogarithm function.
In the limit of vanishing baryon charge, $\mu\to0$, the above result
for $c_R^{(q)}(a,\tilde{\mu})$ reduces to
\beq
c_R^{(q)}(a,\tilde{\mu}=0) =
  \frac{12}{\pi^2}(1-2^{-a})\Gamma(2+a)\zeta(1+a)~.
\label{eq:cRq}
\eeq
In addition, in the special case where $a=1$, $c_R^{(q)}(a,\tilde{\mu})$ becomes a $\tilde{\mu}$-independent constant
\beq
c_R^{(q)}(a=1,\tilde{\mu}) = 2 ~.
\eeq
Numerical values at vanishing chemical potential are listed in table~\ref{tab:cRqg} for various values
of the power $a$ of momentum introduced in
Eq.~(\ref{bulk:discorrection}). Note that both $c_R^{(q)}(a,\tilde{\mu}=0)$ and
$c_R^{(g)}(a)$ increase with $a$ and that $c_R^{(q)}(a,\tilde{\mu}=0)\simeq2c_R^{(g)}(a)$
for large values of $a$. That is, the correction to the
quark loop contribution to screening is twice as large as the
correction to the gluon loop if $a\gg1$.
\begin{table}[hb]
  \begin{tabular}{cccc}
    \hline\hline
    $a$ & 1 & 2 & 3 \\
    \hline
    $c_R^{(q)}(a)$ &  2 & $\frac{54\zeta(3)}{\pi^2}$ &  $\frac{14\pi^2}{5}$ \\
    \hline
    $c_F^{(q)}(a)$ &  3 & $\frac{108\zeta(3)}{\pi^2}$ &  $7\pi^2$ \\
    \hline
    $e^{(q)}(a)$ & 4  & $\frac{13500\zeta(5)}{7\pi^4}$ &
          $\frac{620\pi^2}{49}$ \\
    \hline
    $c_R^{(g)}(a)$ &  2 & $\frac{36\zeta(3)}{\pi^2}$ &  $\frac{8\pi^2}{5}$\\
    \hline
    $c_F^{(g)}(a)$ &  3 & $\frac{72\zeta(3)}{\pi^2}$ &  $ 4\pi^2 $\\
    \hline
    $e^{(g)}(a)$ &  4  & $\frac{1800\zeta(5)}{\pi^4}$ &  $\frac{80\pi^2}{7}$ \\
    \hline
  \end{tabular}
\caption{The numerical coefficients $c_R^{(q,g)}(a)$, $c_F^{(q,g)}(a)$
  and $e^{(q,g)}(a)$ at $\tilde{\mu}=0$ for various values of the power $a$ of momentum
  introduced in the bulk viscous corrections $\delta_{\rm bulk}f$ in
  Eq.~(\ref{bulk:discorrection}).}
\label{tab:cRqg}
\end{table}

Hence, in all we find that bulk viscous corrections ``shift'' the
Debye mass appearing in the retarded self energy by
\bqa
&&\left(2N_c+N_f\left(1+\frac{3 \tilde{\mu}^2}{\pi^2}\right)\right)
\frac{g^2 T^2}{6} \rightarrow \nonumber\\
&&m_{R,D}^2 + \delta m_{R,D}^2 =
\left(2N_c\left(1+c_R^{(g)}(a)\Phi\right)
+N_f\left(1+\frac{3 \tilde{\mu}^2}{\pi^2}\right)
\left(1+c_R^{(q)}(a,\tilde{\mu})\Phi\right)\right)
\frac{g^2 T^2}{6}~. \label{eq:shift_mD_Pi_R}
\eqa

The isotropy of $f_{\rm id}(p) + \delta_{\rm bulk} f(p)$ also implies
that the dependence of the symmetric self energy on energy and
momentum is the same as in equilibrium,
Eqs.~(\ref{eq:Pi_F_eq_quark},\ref{eq:Pi_F_eq_glue}). The correction to
the distribution function written in Eq.~(\ref{bulk:discorrection})
amounts to a shift of the mass scale. For the symmetric self energy to
linear order in $\Phi$ it is\footnote{Note that the mass scale
  obtained from hard thermal loops in equilibrium, $\Phi=0$, is the
  same for the retarded (advanced) and symmetric self energies:
  $m_{R,D}^2=m_{F,D}^2$.}
\bqa
&&\left(2N_c+N_f\left(1+\frac{3 \tilde{\mu}^2}{\pi^2}\right)\right)
\frac{g^2 T^2}{6}\rightarrow \nonumber\\
&&m_{F,D}^2 + \delta m_{F,D}^2 =
\left(2N_c\left(1+c_F^{(g)}(a)\Phi\right)
+N_f\left(1+\frac{3 \tilde{\mu}^2}{\pi^2}\right)
\left(1+c_F^{(q)}(a,\tilde{\mu})\Phi\right)\right)
\frac{g^2 T^2}{6}~.\label{eq:shift_mD_Pi_F}
\eqa
Here,
\beq
c_F^{(q,g)}(a,\tilde{\mu})=\frac{1}{\Phi}
\frac{\int dk k^2 \, \delta_{\rm bulk}f(k) [1 \pm 2f_{\rm id}(k)]}
{\int dk k^2 \, f_{\rm id}(k) [1 \pm f_{\rm id}(k)]} =
\begin{cases}
  \frac{- 3\Gamma(3+a)[{\rm Li}_{(1+a)}(- e^{-\tilde{\mu}})+{\rm Li}_{(1+a)}(- e^{\tilde{\mu}})]}{\pi^2+3 \tilde{\mu}^2} & \text{(fermion)}~, \\
  \frac{3}{\pi^2}\Gamma(3+a)\zeta(1+a) & \text{(boson)}~.
\end{cases}
\eeq
We also list the results for $c_F^{(q)}(a,\tilde{\mu})$ for the two
cases where $\tilde{\mu}=0$ or $a=1$
\bqa c_F^{(q)}(a,\tilde{\mu}=0)
&=& \frac{6}{\pi^2}(1-2^{-a})\Gamma(3+a)\zeta(1+a)\, ,\nonumber
\\ c_F^{(q)}(a=1,\tilde{\mu}) &=& 3~.
\eqa
If $|\Phi|\sim 1$, there are corrections at ${\cal O}(\Phi^2)$ to
$\Pi^{\rm{id}}_F(P)$ which are not negligible. The corresponding
contributions to the Debye mass (divided by its ideal value
$m_{F,D}^2$) are given by
\beq
\begin{cases}
   \Phi^2 \frac{\Gamma(3+2a)}{2(\pi^2+3 \tilde{\mu}^2)}[{\rm Li}_{(2+2a)}(- e^{-\tilde{\mu}})+{\rm Li}_{(2+2a)}(- e^{\tilde{\mu}})-{\rm Li}_{2a}(- e^{-\tilde{\mu}})-{\rm Li}_{2a}(- e^{\tilde{\mu}})] & \text{(fermion)}~, \\
   \Phi^2 \frac{\Gamma(3+2a)}{2\pi^2}[\zeta(2a)-\zeta(2+2a)]\, & \text{(boson)}~.
\end{cases}
\eeq
Recall that at this order in $\Phi$ the validity of the HTL
approximation requires $a>1/2$ and so the $\zeta$-function is well
defined.

%-----------------------------------------------------------------
\section{Expansion about non-thermal fixed point} \label{sec:nonthermalfp}

In this section we compute the temporal component of the gluon self energies using a
non-equilibrium correction inspired by ``anisotropic
hydrodynamics''~\cite{Nopoush:2014pfa}. There the isotropic
non-equilibrium distribution takes the form
\bqa
\label{eq:ex1}
f(p) &=& f_{\rm id}\left(\frac{1}{T}\sqrt{p^2 +
  m^2\left(1+\tilde{\Phi}\right) }\right) \\
&\approx& f_{\rm id}\!(\tilde{p})
  + \frac{m^2 \Phi}{2T\sqrt{p^2 + m^2}} \,
  f_{\rm id}\!(\tilde{p})(1\pm f_{\rm id}\!(\tilde{p}))\, ,
\label{eq:ex2}
\eqa
where $\tilde{p} \equiv \frac{1}{T}\sqrt{p^2 + m^2 }$. Note that here
the scale $m$ is a scalar field expectation value introduced to
skew the ideal distribution from the thermal fixed point (which
would correspond to $m=0$). At weak coupling $m$ does not correspond to
the mass of the quasi-particles which must be obtained from their self
energies, see below.

As already noted in the previous section we must have $|\Phi|\gg g^2$ in
order to be able to compute the gluon self energy at one loop
order. The expansion of $f(p)$ in powers of $\Phi$ furthermore
requires that $|\Phi|\ll1$. This is not a fundamental requirement
though, we expand in powers of $\Phi$ only in order to be able to
provide relatively simple analytic expressions.

Hence, in what follows we consider the bulk viscous correction
\beq
\delta_{\rm bulk} f(p) =
 \frac{m^2 \Phi}{2T \sqrt{p^2 +m^2}} f_{\rm id}\!(\tilde{p}) (1\pm
  f_{\rm id}\!(\tilde{p}))\, .
\eeq
It is straightforward to see that for this correction the gluon loop
contribution to the self energies in the limit $m^2\ll T^2$ is not a
hard thermal loop since it is not dominated by momenta $k\sim T$. We
therefore restrict to $m^2\sim T^2$ (and greater) where the HTL
approximation is applicable.

For $\Phi=0$ the retarded self energy becomes\footnote{In fact,
the self energies corresponding to the distribution function
Eq.~(\ref{eq:ex1}) can be obtained from Eqs.~(\ref{ideal2},
  \ref{eq:Pi00_m_ideal}) for the retarded solution and from
  Eqs.~(\ref{nonideal:re10q}, \ref{nonideal:re10g}) for the symmetric
  solution simply by replacing $\tilde{m}\rightarrow
  \tilde{m}\sqrt{1+\tilde{\Phi}}=\tilde{m}/\sqrt{1+\Phi}$.}
\beq
 \Pi_{R}^{\rm{id}}(P) = N_f\frac{g^2
   T^2}{6}\left(1+\frac{3 \tilde{\mu}^2}{\pi^2}\right)
 f_q(\tilde{m},\tilde{\mu})
 \left(\frac{p_0}{2 p}\ln \frac{p_0+p+i\epsilon}
{p_0-p+i\epsilon} -1\right) ~,\label{ideal2}
\eeq
with
\beq
f_q(\tilde{m},\tilde{\mu})\equiv 2 \left(1+ \frac{- 3\tilde{\mu}\tilde{m}+
   3\tilde{m}\ln[(1+e^{\tilde{\mu}+\tilde{m}})(1+e^{\tilde{\mu}-\tilde{m}})]+
   3[{\rm Li}_2(-e^{\tilde{m}+\tilde{\mu}})+{\rm Li}_2(-e^{\tilde{m}-\tilde{\mu}})]}{\pi^2+3 \tilde{\mu}^2}\right)~,
\eeq
for the fermion loop and
\beq
  \Pi_{R}^{\rm{id}}(P) = 2N_c \frac{g^2 T^2}{6}
  f_g(\tilde{m})
  \left(\frac{p_0}{2 p}\ln \frac{p_0+p+i\epsilon}
{p_0-p+i\epsilon} -1\right)~,   \label{eq:Pi00_m_ideal}
\eeq
with
\beq
f_g(\tilde{m})\equiv \frac{3\tilde{m}^2 +2\pi^2-
    6\tilde{m}\ln(-1+e^{\tilde{m}})-6{\rm Re}[{\rm
      Li}_2(e^{\tilde{m}})]}{\pi^2}~,
\eeq
for the contribution from the boson loop. In the above equations,
$\tilde{m} \equiv \frac{m}{T}$. Evidently, for $m\sim T$ the quasi-particle
masses are still of order $gT$.

The corrections to the self energies of order $\Phi$ are given by
\bqa
 \delta_{{\rm bulk}}    \Pi
_R(P)&=& N_f  \frac{g^2 T^2
}{6} \Phi \, \frac{\tilde{m}^2}{\pi^2}
 \left(\frac{3}{e^{\tilde{m}+\tilde{\mu}}+1} +
 \frac{3}{e^{\tilde{m}-\tilde{\mu}}+1}\right)\left (\frac{p_0}{2 p}\ln
 \frac{p_0+p+i\epsilon}
{p_0-p+i\epsilon} -1\right) \, ,\nonumber \\
 \delta_{{\rm bulk}}    \Pi
_R(P)&=& 2N_c  \frac{g^2
T^2}{6} \Phi \frac{3  \tilde{m}^2/\pi^2}{e^{\tilde{m}}-1}\left
 (\frac{p_0}{2 p}\ln \frac{p_0+p+i\epsilon}
{p_0-p+i\epsilon} -1\right)\, .
\label{nonideal:be2}
\eqa
This corresponds to a shift of the screening mass to
\bqa  \label{eq:mD2_dmD2_mR}
m_{R,D}^2 + \delta m_{R,D}^2 &=&
\left[N_f\left(\left(1+\frac{3
    \tilde{\mu}^2}{\pi^2}\right)f_q(\tilde{m},\tilde{\mu}) + \Phi\,
  \frac{\tilde{m}^2}{\pi^2}
  \left(\frac{3}{e^{\tilde{m}+\tilde{\mu}}+1} +
  \frac{3}{e^{\tilde{m}-\tilde{\mu}}+1}\right)\right)
\right.\nonumber\\
& & + \left.
2N_c \left(f_g(\tilde{m})   + \Phi \frac{3
  \tilde{m}^2/\pi^2}{e^{\tilde{m}}-1}\right)
\right]
 \, \frac{g^2 T^2}{6}~.
\eqa

The symmetric self energy is obtained from the general
expression~(\ref{general:sy})\footnote{The explicit expressions for
  the Fermion contribution to $\Pi_F$ given here apply when
  $m>\mu$. For $m<\mu$ the symmetric self energy could be obtained by
  a numerical evaluation of Eq.~(\ref{general:sy}) with the
  appropriate distribution function.}. For $\Phi=0$,
\begin{equation}
\Pi _{F }^{\rm{id}}(P)=-2\pi iN_f \frac{g^2 T^2
}{ 6 }\frac{T}{p}\frac{6\tilde{m}^2}{\pi^2}
\Theta(p^2-p_0^2)\sum_{n=1}^{\infty} (-1)^{n+1} K_2(\tilde{m} n)
\cosh(\tilde{\mu} n) \, ,
\label{nonideal:re10q}
\end{equation}
for $N_f$ fermion loops and
\begin{equation}
\Pi _{F }^{\rm{id}}(P)=-2 \pi i\, 2 N_c \frac{g^2 T^2
}{ 6 }\frac{T}{p}\frac{3 \tilde{m}^2}{\pi^2} \Theta(p^2-p_0^2)\sum_{n=1}^{\infty}  K_2(\tilde{m} n)\, ,
\label{nonideal:re10g}
\end{equation}
for the boson loops. As before $\tilde{m} \equiv \frac{m}{T}$ is
assumed to be of order $1$ or greater. $K_n(z)$ denotes the modified
Bessel function of the second kind.

The corresponding bulk-viscous corrections are given by
\bqa
\delta_{\rm {bulk}}    \Pi_F(P) &=& -2\pi iN_f \frac{g^2 T^2}{6}
\frac{T}{p} \frac{3 \Phi \tilde{m}^3}{\pi^2}
\Theta(p^2-p_0^2)
\sum_{n=1}^{\infty} n (-1)^{n+1} K_1(\tilde{m} n) \cosh(\tilde{\mu} n)\, , \nonumber \\
\delta_{{\rm bulk}}    \Pi_F(P) &=&  -2\pi i \, 2N_c \frac{g^2 T^2}{6}
\frac{T}{p}\frac{3 \Phi \tilde{m}^3}{2 \pi^2}
\Theta(p^2-p_0^2)\sum_{n=1}^{\infty} n K_1(\tilde{m} n)\, .
\label{nonideal:be2F}
\eqa
Hence, the corrections have a different dependence on $\tilde{m} \equiv
\frac{m}{T}$ than the ``ideal'' contributions.

For the symmetric self energy, the bulk viscous correction also
corresponds to a shift of the mass scale to
\bqa  \label{eq:mD2_dmD2_mF}
m_{F,D}^2 + \delta m_{F,D}^2 &=&
\left[N_f \frac{6\tilde{m}^2}{\pi^2} \sum_{n=1}^{\infty}
  (-1)^{n+1} \cosh(\tilde{\mu} n)\bigg( K_2(\tilde{m} n) + \Phi\frac{\tilde{m}}{2}
 n K_1(\tilde{m} n)\bigg)
\right.\nonumber\\
& & + \left.
2N_c \frac{3 \tilde{m}^2}{\pi^2}\sum_{n=1}^{\infty}\bigg(
K_2(\tilde{m} n) + \Phi \frac{ \tilde{m}}{2}
 n K_1(\tilde{m} n)\bigg)
\right]\times \frac{g^2 T^2}{6}~.
\eqa
Notice that in general the Debye mass obtained from the ideal
distribution is different for the retarded (advanced) and symmetric
solutions. Only at the thermal fixed point, i.e.\ in thermal
equilibrium with $m=0$ as studied in the previous section, do we have
$m_{R,D}^2=m_{F,D}^2=(2N_c+N_f(1+\frac{3 \tilde{\mu}^2}{\pi^2})) \frac{g^2 T^2}{6}$.

As before, we also give the ${\cal O}(\Phi^2)$ contributions to the
mass $m_{F,D}^2$:
\beq
\begin{cases}
   \Phi^2 N_f \frac{g^2 T^2}{6}\frac{\tilde{m}^4 }{4\pi^2}\sum_{n=1}^{\infty}
  (-1)^n n (n+1)(n+2) {\cal G} (\tilde{m},n)\cosh[\tilde{\mu}(n+1)]~& \text{(fermion)}~ , \\
   \Phi^2 2N_c \frac{g^2 T^2}{6}\frac{\tilde{m}^4 }{8\pi^2}\sum_{n=1}^{\infty}
  n (n+1)(n+2) {\cal G} (\tilde{m},n)~ & \text{(boson)}~.
\end{cases}
\eeq
In the above expressions, the function ${\cal G} (\tilde{m},n)$ is defined as
\begin{equation}
{\cal G} (\tilde{m},n)=\int_{\tilde{m}}^{\infty}\frac{\sqrt{t^2-\tilde{m}^2}}{t}e^{-(n+1)t}dt\, .
\end{equation}

%-----------------------------------------------------------------
\section{Bulk viscous correction to the gluon propagator}

In this section we compute the HTL resummed propagator for
longitudinal gluons. We employ Coulomb gauge with gauge parameter set to zero.

The system under consideration is still isotropic after including the
bulk viscous correction to the distribution function. For such a
system in Coulomb gauge, the temporal component of the resummed
propagator is independent of the spatial components of the self energy
and bare propagator and it can be determined through the following
Schwinger-Dyson equation \footnote{Since only temporal components
  appear for all the propagators and self energies, we omit the
  superscript ``$00$" in the following.}
\beq
{\tilde{D}^*}_{R}(P)= D_R(P)+ D_R(P)\,\tilde{\Pi}_{R}(P)\,
{\tilde{D}^*}_{R}(P)\, ,  \label{eq:SD_D_R}
\eeq
with
\beq
\label{2a10}
D_R(P)  =  D_A(P)= \frac{1}{p^2}\, ,
\eeq
the (temporal component of) the non-resummed real-time propagators for
gluons. Here, a superscript star on propagators indicates
 a resummed propagator. We define $\tilde{\Pi}_{R/A/F}\equiv \Pi^{\rm {id}}_{R/A/F}+ \delta_{\rm{bulk}}\Pi_{R/A/F}$
and the same definition holds for ${\tilde{D}^*}_{R/A/F}$.
Eq.~(\ref{eq:SD_D_R}) is solved by
\beq \label{probc}
{\tilde{D}^*}_{R}(P) = \frac{1} {p^2 -
  \tilde{\Pi}_{R}(P)} = \frac{1} {p^2 - \left(m_{R,D}^2 +
  \delta m_{R,D}^2\right)\left (\frac{p_0}{2 p}\ln
  \frac{p_0+p+i\epsilon} {p_0-p+i\epsilon} -1\right)}~.
\eeq
This propagator applies for momenta of order $\sqrt{m_{R,D}^2 +\delta m_{R,D}^2}$ (or greater). If $\delta_{\rm {bulk}}\Pi_{R}\ll \Pi^{\rm {id}}_{R}$ or
$|\Phi|\ll1$ then bulk viscous corrections are small. On the other hand,
for $|\Phi|\sim 1$ the propagator resums insertions of $\delta_{\rm {bulk}}\Pi_{R}$
into each hard thermal loop. For $|\Phi|=0$, one has the well known result
\beq
{\tilde{D}^*}_{R}(P)=\frac{1}{p^2-m_{R,D}^2\left(\frac{p_0}{2 p}\ln
\frac{p_0+p+i\epsilon} {p_0-p+i\epsilon} -1\right)}\, ,\label{3b17}
\eeq
where $m_{R,D}^2=(2N_c+N_f)\, g^2T^2/6$ for massless particles in thermal
equilibrium. For the model from Sec.~\ref{sec:nonthermalfp} which
expands about a non-thermal distribution function the screening mass
depends on the scale $m$ and is given in Eq.~(\ref{eq:mD2_dmD2_mR}).  The
advanced propagator is obtained by the replacement $i\epsilon\to
-i\epsilon$.

The symmetric (time ordered) resummed propagator is obtained from
\begin{eqnarray}\label{df}
{\tilde{D}^*}_{F}(P)&=&  (1+2\tilde{f}(p_0))\, \mbox{sgn}(p_0)\,
[{\tilde{D}^*}_{R}(P)-{\tilde{D}^*}_{A}(P)]\,\\ \nonumber
&+&{\tilde{D}^*}_{R}(P)\{\tilde{\Pi}_F(P)-[1+2 \tilde{f}(p_0)]\, \mbox{sgn}(p_0)\, [\tilde{\Pi} _R(P)-\tilde{\Pi}
_A(P)]\}{\tilde{D}^*}_{A}(P)\, ,
\end{eqnarray}
where $\tilde{f}(p_0)\equiv f_{\rm{id}}(p_0)+\delta_{\rm{bulk}}f(p_0)$.

In thermal equilibrium the KMS relation implies that
$\Pi^{\rm{id}}_F(P)=[1+2 f_{\rm id}(p_0)]\, \mbox{sgn}(p_0)\,
[\Pi^{\rm{id}}_R(P)-\Pi^{\rm{id}}_A(P)]$ and so the second line on the r.h.s.\ of the
previous equation vanishes.

Our model with $m\sim T$ involves an ideal distribution corresponding
to a non-thermal fixed point. In an unbroken theory such as QED or QCD
the gauge bosons are massless and no mass appears in the bare
propagators $D(P)$, c.f.\ Eqs.~(\ref{2a10}). Indeed, the gluon
self-energies derived in previous sections have been computed using
massless propagators. Rather, the scalar mass-like scale $m$ is merely
a parameter which distorts the distribution function $f_{\rm id}(
p_0)$ from the thermal fixed point and so the KMS relation for
$\Pi^{\rm{id}}_F(P)$ does not apply. Hence, the second line in Eq.~(\ref{df})
does not vanish in the ``ideal limit'' $\Phi\to0$. This can be checked
easily by taking the limit $p_0\to 0$ of the expressions for $\Pi^{\rm{id}}_{R/A/F}$
 given in Sec.~\ref{sec:nonthermalfp}.

Using the identity
\beq
{\tilde{D}^*
}_{R}(P)-{\tilde{D}^*}_{A}(P)
={\tilde{D}^*
}_{R}(P)\,[\tilde{\Pi}_{R}(P)- \tilde{\Pi}_{A}(P)]  \,
{\tilde{D}^*}_{A}(P)\, ,
\eeq
which follows from Eq.~(\ref{probc}) and an analogous expression for
the resummed advanced propagator, Eq.~(\ref{df}) can be simplified to
\beq
{\tilde{D}^*}_{F}(P)=  {\tilde{D}^*}_{R}(P)\,\tilde{\Pi} _{F}(P)  \,
{\tilde{D}^*}_{A}(P) \, . \label{dft}
\eeq

Eqs.~(\ref{probc}) and (\ref{dft}) are the main results of this
section. They are applicable for both models introduced above.

%-----------------------------------------------------------------
\section{Static potential}

In this section we apply the results obtained above to the QCD static
potential at finite temperature. We define the static potential due to
one gluon exchange through the Fourier transform of the physical ``11"
Schwinger-Keldysh component of the (longitudinal) gluon propagator in
the static limit~\cite{Dumitru:2009fy}:
\begin{eqnarray}
V({\bf{r}}) &=& (ig)^2 C_F\int \frac{d^3{\bf{p}}}{(2\pi)^3} \,
\left(e^{i{\bf{p \cdot r}}}-1\right)\, \left({\tilde{D}^*}(p_0=0,
  \bf{p})\right)_{11} \nonumber\\
&=& -g^2 C_F\int \frac{d^3{\bf{p}}}{(2\pi)^3} \, \left(e^{i{\bf{p \cdot
r}}}-1\right)\,
\frac{1}{2}\left({\tilde{D}^*}_R+{\tilde{D}^*}_A+{\tilde{D}^*}_F\right)
\nonumber \\
&=& -g^2 C_F\int \frac{d^3{\bf{p}}}{(2\pi)^3} \, \left(e^{i{\bf{p \cdot
r}}}-1\right)\,
\frac{1}{2}\left({\tilde{D}^*}_R+{\tilde{D}^*}_A\right)\nonumber\\
&& -g^2 C_F\int \frac{d^3{\bf{p}}}{(2\pi)^3} \, \left(e^{i{\bf{p \cdot
r}}}-1\right)\, \frac{1}{2}{\tilde{D}^*}_F ~. \label{41}
\end{eqnarray}
We have taken the sources in the fundamental representation and
subtracted an $r$-independent (self-energy) contribution.  In the
static limit, $\frac{1}{2}\left({\tilde{D}^*}_R + {\tilde{D}^*}_A\right) =
{\tilde{D}^*}_R = {\tilde{D}^*}_A$. The Fourier transform of this quantity
gives the real part of the (screened) potential while its imaginary
part, describing Landau damping~\cite{Laine:2006ns}, comes from the
Fourier transform of the symmetric propagator.

In the limit $p_0\to0$ the retarded or advanced self energies equal
(minus) the square of the screening mass and so the Fourier transform
of Eq.~(\ref{probc}) gives
\beq \label{eq:Re_V}
\mathrm{Re}\, V(r) = - \frac{g^2 C_F}{4\pi r}\,
e^{-r\, \sqrt{m_{R,D}^2 + \delta m_{R,D}^2}}
+ 2 F_Q(m_{R,D})~.
\eeq
Expressions for $m_{R,D}^2 + \delta m_{R,D}^2$ have been given in
Eqs.~(\ref{eq:shift_mD_Pi_R},\ref{eq:mD2_dmD2_mR}) above.
This potential applies to distance scales of order $1/m_{R,D}$ or less,
where $m_{R,D}=\sqrt{m_{R,D}^2 + \delta m_{R,D}^2}$. Also, in
Eq.~(\ref{eq:Re_V}) we have restored the $r$-independent but
$T$-dependent free energy contribution
\beq
2 F_Q = g^2 C_F\int \frac{d^3{\bf{p}}}{(2\pi)^3}
\left({\tilde{D}^*}_R-{D}_R\right)
= - \frac{g^2 C_F}{4\pi}\, \sqrt{m_{R,D}^2 + \delta m_{R,D}^2}~.
\eeq

The imaginary part of the potential is given by\footnote{The
  expression for $\mathrm{Im}\, V(r)$ at the thermal fixed point
  without non-equilibrium corrections was first derived in
  Refs.~\cite{Laine:2006ns} and generalized to anisotropic
  shear-viscous corrections in Ref.~\cite{Dumitru:2009fy}.
  For recent lattice measurements of the real and imaginary
parts of the static potential in equilibrium, see
Ref.~\cite{Burnier:2014ssa}.}
\begin{equation}
\mathrm{Im}\, V(r) = - \frac{ g^2 C_F T}{4 \pi } \frac{m_{F,D}^2 +
  \delta m_{F,D}^2}{m_{R,D}^2 + \delta m_{R,D}^2} \,
\phi(\hat{r})\, , \label{46}
\end{equation}
where
\begin{equation}
 \phi(\hat{r})= 2\int_0^{\infty}dz \frac{
z}{(z^2+1)^2} \left[1-\frac{\sin(z\, \hat{r})}{z\,
\hat{r}}\right]\, , \label{47}
\end{equation}
with $\hat{r} \equiv r  \sqrt{m_{R,D}^2 + \delta m_{R,D}^2}$.
The expressions for $m_{F,D}^2 + \delta m_{F,D}^2$ have been given in
Eqs.~(\ref{eq:shift_mD_Pi_F},\ref{eq:mD2_dmD2_mF}) above. For small
$\hat r$ the function $\phi(\hat{r})$ is proportional to $\hat{r}^2\,{\rm {ln}}\hat{r}$.

From the above results, we can conclude that since the bulk viscous
corrections are isotropic the real part of the potential has the same
structure as in the ideal case with $m_{R,D}^2$ replaced by $m_{R,D}^2 +
\delta m_{R,D}^2$. The imaginary part of the potential is multiplied
by a factor $\frac{m_{F,D}^2 + \delta m_{F,D}^2}{m_{R,D}^2 + \delta
  m_{R,D}^2}$ which equals $1$ in thermal equilibrium.

%-----------------------------------------------------------------
\section{Application to heavy-ion collisions: Landau matching}
\label{sec:LandauMatch}

There are two sources of corrections when comparing an ideal to a
viscous thermal medium. There are, of course, corrections to the
hydrodynamic evolution equations as well as to the initial
conditions\footnote{For applications to heavy-ion collisions the
  initial condition for viscous hydrodynamics, e.g.\ the initial
  temperature etc., is adapted such as to reproduce the {\em measured}
final state of the collision. For example, the observed charged hadron
multiplicity constrains the entropy in the final state and so
on.}. Second, there are {\em explicit} corrections to observables such
as the heavy quark potential considered above.

In order not to mix these corrections one matches the 00-components of
the energy momentum tensors of the ideal and viscous fluids,
respectively, in the local rest frames. That is, in the local rest
frame the non-equilibrium corrections should not contribute to the
energy density. For ``anisotropic hydrodynamics'' the matching is done
in Ref.~\cite{Nopoush:2014pfa}, here we focus on the model introduced
in Sec.~\ref{sec:thermalfp} where the ideal distribution corresponds
to the thermal fixed point. We also set $\mu=0$ for simplicity.

To match the energy densities we shift the temperature of the viscous
medium to $T'$ which is determined from\footnote{Notice that for the
  quark contribution, there is a pre-factor $2N_f$ counting the number
  of quarks. For the gluon contribution, the factor is $N_c^2-1$.}
\beq
\int \frac{d^3k}{(2\pi)^3} E_k f_{\rm id}(k;T) =
\int \frac{d^3k}{(2\pi)^3} E_k \tilde{f}(k;T')~.
\eeq
With the bulk viscous correction from Eq.~(\ref{bulk:discorrection})
this leads to
\beq
T^4 = {T'}^4 \left[ 1+ \Phi
\frac{2(N_c^2-1)e^{(g)}(a)+4 N_f\frac{7}{8}e^{(q)}(a)}{2(N_c^2-1) +
  4 N_f\frac{7}{8}} \right]~.
\eeq
To invert this relation for simplicity we now assume that $\Phi$ is a
small parameter so that to linear order in $\Phi$
\beq
T' \simeq T \left[ 1- \frac{1}{4}\Phi
\frac{2(N_c^2-1)e^{(g)}(a)+4N_f\frac{7}{8}e^{(q)}(a)}
{2(N_c^2-1)+4N_f\frac{7}{8}}
\right]~.  \label{eq:Tprime_T}
\eeq
The numbers $e^{(g)}(a)$ and $e^{(q)}(a)$ are defined as follows:
\bqa
e^{(g)}(a) &=& \frac{1}{\Phi\, {T'}^4} \frac{30}{\pi^2} \int
\frac{dk}{2\pi^2} k^3\,
\delta_{\rm bulk} f(k;T')~,\\
e^{(q)}(a) &=& \frac{1}{\Phi\, {T'}^4} \frac{8\cdot30}{7\pi^2} \int
\frac{dk}{2\pi^2} k^3\,
\delta_{\rm bulk} f(k;T')~,
\eqa
and have been listed (for $a=1$, 2, 3) in table~\ref{tab:cRqg}.

The temperature $T$ which appears in the gluon self energies and
propagators, and in the static potential, should now be replaced by
$T'$ as given in Eq.~(\ref{eq:Tprime_T}).

As an example, consider the case $a=1$ so that $e^{(g)}(1)=e^{(q)}(1)=
4$. We then obtain ${T'}^2\simeq T^2(1-2\Phi)$ if $|\Phi|\ll1$. Since
$c_R^{(g)}(1) = c_R^{(q)}(1)=2$ in all we find that the ``shift'' of
the Debye mass appearing in the retarded self energy,
Eq.~(\ref{eq:shift_mD_Pi_R}), cancels. Hence, in this case there are
no {\em explicit} bulk viscous corrections to the retarded gluon self
energy. It is only affected by the implicit change of initial
conditions and hydrodynamic solution in the presence of a
non-vanishing bulk viscosity. On the other hand, for the symmetric
self energy $c_F^{(g)}(1) = c_F^{(q)}(1)=3$ and so there is an
explicit correction
\beq
(2N_c+N_f) \frac{g^2 T^2}{6} ~~\rightarrow~~
\left(2N_c\left(1+\Phi\right)
+N_f\left(1+\Phi\right)\right)
\frac{g^2 T^2}{6}~,
\eeq
even after Landau matching has been performed. Since $|\Phi|\sim\zeta$,
the correction inherits the dynamical critical scaling of the bulk
viscosity in the vicinity of a second order critical point.

Finally, we write the correction to the pressure which for the model
from Sec.~\ref{sec:thermalfp} is given by
\beq
\frac{\delta_{\rm bulk}p}{p_{\rm id}} = \Phi\,
\left(\frac{T'}{T}\right)^4\,
\frac{2(N_c^2-1)e^{(g)}(a)+4 N_f\frac{7}{8}e^{(q)}(a)}
{2(N_c^2-1)+4 N_f\frac{7}{8}}~.
\eeq
Generically one expects a negative bulk pressure, so $\Phi<0$, unless
its sign is reversed by shear-bulk coupling~\cite{BulkShearKinetic}.

\section{Summary and Discussion} \label{sec:summary}

Non-equilibrium corrections to the distribution functions of quarks
and gluons in a hot and dense QCD medium result in corrections to
``hard thermal loops'' (HTL) which define Debye screening and Landau
damping. In this paper we have considered two different forms of
bulk-viscous corrections to ideal distributions corresponding to
either thermal distributions or to a non-thermal fixed point obtained
by introducing a non-vanishing scalar field. We find that the gluon
hard thermal loop is dominated by hard momenta provided that the
bulk-viscous corrections are sufficiently suppressed (relative to Bose
enhanced behavior $f_B(k) (1+f_B(k))\sim (T/k)^2$) for momenta $k\ll
T$; for the quark loop this is ensured by Pauli blocking.

Our main result is that isotropic bulk-viscous corrections shift the
screening and damping mass scales which appear in the
retarded/advanced versus the symmetric gluon HTL self energies. The
shift is different for the two types of self energies. For example,
bulk-viscous corrections to the thermal fixed point lead to the replacement
\bqa
\left[2N_c+N_f\left(1+\frac{3 \tilde{\mu}^2}{\pi^2}\right)\right]
\frac{g^2 T^2}{6} &\rightarrow& \nonumber\\
m_{R,D}^2 + \delta m_{R,D}^2 &=&
\left[2N_c\left(1+c_R^{(g)}\Phi\right)
+N_f\left(1+\frac{3 \tilde{\mu}^2}{\pi^2}\right)
\left(1+c_R^{(q)}\Phi\right)\right]
\frac{g^2 T^2}{6} \label{eq:summary_shift_mD_Pi_R}~,
\eqa
in the retarded self energy (screening mass), and to
\bqa
\left[2N_c+N_f\left(1+\frac{3 \tilde{\mu}^2}{\pi^2}\right)\right]
\frac{g^2 T^2}{6}&\rightarrow& \nonumber\\
m_{F,D}^2 + \delta m_{F,D}^2 &=&
\left[2N_c\left(1+c_F^{(g)}\Phi\right)
+N_f\left(1+\frac{3 \tilde{\mu}^2}{\pi^2}\right)
\left(1+c_F^{(q)}\Phi\right)\right]
\frac{g^2 T^2}{6}\label{eq:summary_shift_mD_Pi_F}~,
\eqa
in the symmetric self energy (to linear order in $\Phi$). Here,
$\tilde{\mu}$ is the quark-chemical potential divided by temperature
$T$ and $\Phi$ is proportional to the bulk pressure divided by the
ideal pressure. The $c_{R/F}^{(q,g)}$ are coefficients which we
computed (also see Sec.~\ref{sec:LandauMatch} on how to determine the
temperature of the non-equilibrium system).

In the absence of strong shear-bulk coupling, generically $\Phi<0$
which implies reduced screening and damping scales. In particular, at
a second order critical point it is expected that, up to finite time
and finite size effects, the bulk viscosity diverges $\sim\xi^z$ as a
power (dynamical critical exponent) of the correlation length. This
reflects the coupling of the hydrodynamical modes, i.e.\ of
fluctuations of conserved currents, to those of the light, slow order
parameter. Our analysis shows that in general the screening and
damping mass scales are also sensitive to such increase of the bulk
viscosity.  This could reflect, for example, in the properties of
quarkonium bound states measured in heavy-ion
collisions~\cite{QQbar_HIC}.

Non-equilibrium bulk-viscous corrections also affect the dynamics of
high occupancy soft fields~\cite{Blaizot:1993zk} which is given by the
classical Yang-Mills equations,
\beq \label{eq:DFmunu}
D_\mu F^{\mu\nu} = 2 \left(m_{R,D}^2 + \delta m_{R,D}^2\right)
\int\frac{d^3v}{4\pi} v^\nu w(\mathbf{x},\mathbf{v})~.
\eeq
$w(\mathbf{x},\mathbf{v})$ describes color charge fluctuations
due to hard particles. It satisfies
\beq
v^\mu D_\mu w(\mathbf{x},\mathbf{v}) = \mathbf{v}\cdot\mathbf{E}~.
\eeq
Eq.~(\ref{eq:DFmunu}) involves the shifted screening mass squared
$m_{R,D}^2 + \delta m_{R,D}^2$. If negative this would lead to
instabilities of the soft gauge fields.

\section*{Acknowledgements}
Q.D. and Y.G.\ gratefully acknowledge supports by the NSFC of China under Project No. 11665008, by
Natural Science Foundation of Guangxi Province of China under Project No. 2016GXNSFFA380014 and by the ``Hundred
Talents Plan" of Guangxi Province of China. A.D.\ gratefully acknowledges support by the DOE
Office of Nuclear Physics through Grant No.\ DE-FG02-09ER41620; and
from The City University of New York through the PSC-CUNY Research
grant 69362-0047. M.S. was supported by the U.S. Department of Energy, Office of Science, Office of
Nuclear Physics under Award No. DE-SC0013470.

\appendix

\section{Gluon Self Energies in the Real Time Formalism}

In this appendix, we employ the real time formalism of thermal field
theory to compute the gluon self energies at one-loop order within the
Hard Thermal Loop (HTL) approximation. In finite temperature field
theory, the real time formalism is more appropriate when dealing with
a non-equilibrium situation. The corresponding results at vanishing
chemical potential $\mu$ have already been obtained before; see, for
example,
Ref.\cite{Dumitru:2009fy,Mrowczynski:2000ed,Carrington:1997sq}. Here,
we recompute the gluon self energies at finite chemical potential by
keeping track of the quark distribution function $f^+_F({\bf k})$ and
the anti quark distribution function $f^-_F({\bf k})$ explicitly
during the calculation.

In the Keldysh representation, only the symmetric component of the bare
fermion propagators depends on the chemical potential. It reads
\begin{equation}
S_{F}(K)=-2\pi i K\!\!\!\!/\,[1-2(\Theta(k_{0})f^+_F+\Theta(-k_{0})f^-_F)]\,\delta(K^{2})~.
\end{equation}
The retarded/advanced components are given by
\bqa
S_{R}(K)&=&\frac{K\!\!\!\!/}{K^{2}+i \, {\rm sgn}(k_{0})\,\epsilon}\,
,\nonumber \\ S_{A}(K)&=& \frac{K\!\!\!\!/}{K^{2}-i \, {\rm
    sgn}(k_{0})\,\epsilon}\,,
\eqa
where ${\rm sgn}(x)$ is the sign function. We neglect the fermion mass
in our calculation.

%In addition, for the bare boson propagators, we have
%\bqa
%D_{R}(K)&=&\frac{1}{K^{2}+i \, {\rm sgn}(k_{0})\,\epsilon}\, ,\nonumber \\
%D_{A}(K)&=&\frac{1}{K^{2}-i  \, {\rm sgn}(k_{0})\,\epsilon}\, ,\nonumber \\
%D_{F}(K)&=&-2\pi i(1+2f_{B})\delta (K^{2})\, .
%\eqa

We only need to consider the Feynman diagram with a quark loop, since
it is the only one at one-loop order that depends on the chemical
potential. For the retarded/advanced gluon self-energy, the temporal component can
be expressed as
\begin{eqnarray}
  \Pi
_R (P)&=&-i N_f g^2\int \frac{d^4K}{(2\pi )^4} (q_0k_0+{\bf
q}\cdot {\bf k}) \biggl [\tilde \Delta _F(Q)\tilde \Delta
_R(K)+\tilde \Delta _A(Q)
\tilde \Delta _F(K)\nonumber \\
&& +\tilde \Delta _A(Q)\tilde \Delta _A(K)+\tilde \Delta
_R(Q)\tilde \Delta _R(K)\biggr ]\, . \label{3b9}
\end{eqnarray}
Here, $S _{R/A/F}(K)\equiv K\!\!\!\!/ \tilde \Delta _{R/A/F}(K)$ and
$Q=K-P$. The last two terms of the integrand vanish after integration
over $k_0$. At vanishing chemical potential, one can show that the first two
terms contribute equally to the final result by the
substitution $K \rightarrow - K+P$\footnote{For a non-equilibrium
  distribution this requires that $f({\bf k})=f(-{\bf k})$.}.  However,
this is no longer true when $\mu \neq 0$. With the same replacement,
we can recombine the contributions from the first two terms and
finally arrive at
\begin{eqnarray}
  \Pi
_R(P)&=&2 \pi N_f g^2 \int \frac{k d k d
\Omega}{(2\pi )^4} (f^+_F({\bf k})+f^-_F({\bf k}))
\biggl [ \frac{2 k^2 - p_0 k-{\bf k}\cdot {\bf p}}{P^2-2 k p_0+2{\bf k}\cdot {\bf p}-i\, \mbox{sgn}(k - p_0)\epsilon}\nonumber \\
&& +\frac{2 k^2 + p_0 k-{\bf k}\cdot {\bf p}}{P^2+2 k p_0+2{\bf
k}\cdot {\bf p}-i\, \mbox{sgn}(-k-p_0)\epsilon}\biggr ]\, .
\end{eqnarray}
The remainder of the calculation is very similar to the $\mu=0$
case. In the HTL approximation, the leading contribution is given by
Eq.~(\ref{general:re}).

Furthermore, for the temporal component of the symmetric gluon self-energy, we have
\begin{eqnarray}
 \Pi _F(P)&=&-i N_f g^2\int
\frac{d^4K}{(2 \pi )^4}(q_0k_0+{\bf q}\cdot {\bf k})\biggl
[\tilde \Delta _F(Q)\tilde \Delta _F(K)-(\tilde \Delta _R(Q)
-\tilde \Delta _A(Q))\nonumber \\
&&\times (\tilde \Delta _R(K)-\tilde \Delta _A(K))\biggr ]\, .
\label{3b19}
\end{eqnarray}
Using the HTL approximation, a straightforward calculation leads to
\begin{eqnarray}
 \Pi _F(P)&=&4 i N_f g^2 \pi^2\int
\frac{k^2dkd\Omega}{(2 \pi )^4}\frac{2}{p}\biggl[f^+_F({\bf k})(f^+_F({\bf k})-1)
\delta({\hat{\bf k}}\cdot {\hat{\bf
    p}}-\frac{p_0}{p})\nonumber \\
    &+&f^-_F({\bf k})(f^-_F({\bf k})-1)\delta({\hat{\bf k}}\cdot {\hat{\bf
    p}}+\frac{p_0}{p})\biggr ] \, .
\end{eqnarray}
If the distribution function only depends on the modulus of the
momentum $\mathbf{k}$, we can simplify this further. Notice that
although the arguments of the two delta functions are different that
nevertheless they give the same contribution after integrating over
${\rm d}\Omega$. Finally, we arrive at
\begin{eqnarray}
\label{pifiso}
 \Pi _F(P)=4 i N_f g^2 \pi^2\int
\frac{k^2dk}{(2 \pi )^3}\frac{2}{p}\biggl[f^+_F(k)(f^+_F(k)-1)
+f^-_F(k)(f^-_F(k)-1)\biggr ]\Theta(p^2-p_0^2)\, .
\end{eqnarray}
For thermal equilibrium distributions we can rewrite the above equation as
\begin{eqnarray}
%\label{pifiso}
 \Pi _F(P)=4 i N_f g^2 T \pi^2\int
\frac{k^2dk}{(2 \pi )^3}\frac{2}{p}\biggl(\frac{d n^+_F(k)}{d k}+\frac{d n^-_F(k)}{d k}\biggl)\Theta(p^2-p_0^2)\, .
\end{eqnarray}
%

%-----------------------------------------------------------------------


\begin{thebibliography}{99}

\bibitem{Weldon:1982aq}
  H.~A.~Weldon,
  %``Covariant Calculations at Finite Temperature: The Relativistic Plasma,''
  Phys.\ Rev.\ D {\bf 26}, 1394 (1982).
  %doi:10.1103/PhysRevD.26.1394
  %%CITATION = doi:10.1103/PhysRevD.26.1394;%%
  %708 citations counted in INSPIRE as of 24 Oct 2016

\bibitem{Braaten:1989mz}
  E.~Braaten and R.~D.~Pisarski,
  %``Soft Amplitudes in Hot Gauge Theories: A General Analysis,''
  Nucl.\ Phys.\ B {\bf 337}, 569 (1990).
  %doi:10.1016/0550-3213(90)90508-B
  %%CITATION = doi:10.1016/0550-3213(90)90508-B;%%
  %1158 citations counted in INSPIRE as of 24 Oct 2016

\bibitem{Frenkel:1989br}
  J.~Frenkel and J.~C.~Taylor,
  %``High Temperature Limit of Thermal QCD,''
  Nucl.\ Phys.\ B {\bf 334}, 199 (1990).
  %doi:10.1016/0550-3213(90)90661-V
  %%CITATION = doi:10.1016/0550-3213(90)90661-V;%%
  %478 citations counted in INSPIRE as of 24 Oct 2016

\bibitem{Braaten:1991gm}
  E.~Braaten and R.~D.~Pisarski,
  %``Simple effective Lagrangian for hard thermal loops,''
  Phys.\ Rev.\ D {\bf 45}, no. 6, R1827 (1992).
  %doi:10.1103/PhysRevD.45.R1827
  %%CITATION = doi:10.1103/PhysRevD.45.R1827;%%
  %326 citations counted in INSPIRE as of 24 Oct 2016

\bibitem{Haque:2014rua}
  N.~Haque, A.~Bandyopadhyay, J.~O.~Andersen, M.~G.~Mustafa, M.~Strickland and N.~Su,
  %``Three-loop HTLpt thermodynamics at finite temperature and chemical potential,''
  JHEP {\bf 1405}, 027 (2014).
  %doi:10.1007/JHEP05(2014)027
  %[arXiv:1402.6907 [hep-ph]].
  %%CITATION = doi:10.1007/JHEP05(2014)027;%%
  %69 citations counted in INSPIRE as of 24 Oct 2016

\bibitem{Andersen:2011sf}
  J.~O.~Andersen, L.~E.~Leganger, M.~Strickland and N.~Su,
  %``Three-loop HTL QCD thermodynamics,''
  JHEP {\bf 1108}, 053 (2011).
  %doi:10.1007/JHEP08(2011)053
  %[arXiv:1103.2528 [hep-ph]].
  %%CITATION = doi:10.1007/JHEP08(2011)053;%%
  %79 citations counted in INSPIRE as of 24 Oct 2016

\bibitem{Andersen:1999fw}
  J.~O.~Andersen, E.~Braaten and M.~Strickland,
  %``Hard thermal loop resummation of the free energy of a hot gluon plasma,''
  Phys.\ Rev.\ Lett.\  {\bf 83}, 2139 (1999).
  %doi:10.1103/PhysRevLett.83.2139
  %[hep-ph/9902327].
  %%CITATION = doi:10.1103/PhysRevLett.83.2139;%%
  %183 citations counted in INSPIRE as of 24 Oct 2016

\bibitem{Laine:2006ns}
M.~Laine, O.~Philipsen, P.~Romatschke and M.~Tassler,
%``Real-time static potential in hot QCD,''
JHEP {\bf 0703}, 054 (2007);\\
%doi:10.1088/1126-6708/2007/03/054
%[hep-ph/0611300].
%%CITATION = doi:10.1088/1126-6708/2007/03/054;%%
N.~Brambilla, J.~Ghiglieri, A.~Vairo and P.~Petreczky,
%``Static quark-antiquark pairs at finite temperature,''
Phys.\ Rev.\ D {\bf 78}, 014017 (2008);\\
%doi:10.1103/PhysRevD.78.014017
%[arXiv:0804.0993 [hep-ph]].
%%CITATION = doi:10.1103/PhysRevD.78.014017;%%
M.~A.~Escobedo and J.~Soto,
%``Non-relativistic bound states at finite temperature (I): The Hydrogen atom,''
Phys.\ Rev.\ A {\bf 78}, 032520 (2008).
%doi:10.1103/PhysRevA.78.032520
%[arXiv:0804.0691 [hep-ph]].
%%CITATION = doi:10.1103/PhysRevA.78.032520;%%

\bibitem{Strickland:2013uga}
  M.~Strickland,
  %``Thermalization and isotropization in heavy-ion collisions,''
  Pramana {\bf 84}, no. 5, 671 (2015).
  %doi:10.1007/s12043-015-0972-1
  %[arXiv:1312.2285 [hep-ph]].
  %%CITATION = doi:10.1007/s12043-015-0972-1;%%
  %25 citations counted in INSPIRE as of 11 Nov 2016

\bibitem{Strickland:2014pga}
  M.~Strickland,
  %``Anisotropic Hydrodynamics: Three lectures,''
  Acta Phys.\ Polon.\ B {\bf 45}, no. 12, 2355 (2014).
  %doi:10.5506/APhysPolB.45.2355
  %[arXiv:1410.5786 [nucl-th]].
  %%CITATION = doi:10.5506/APhysPolB.45.2355;%%
  %37 citations counted in INSPIRE as of 11 Nov 2016

\bibitem{Dumitru:2007hy}
A.~Dumitru, Y.~Guo and M.~Strickland,
%``The Heavy-quark potential in an anisotropic (viscous) plasma,''
Phys.\ Lett.\ B {\bf 662}, 37 (2008).
%doi:10.1016/j.physletb.2008.02.048
%[arXiv:0711.4722 [hep-ph]].
%%CITATION = doi:10.1016/j.physletb.2008.02.048;%%

\bibitem{Dumitru:2009fy}
A.~Dumitru, Y.~Guo and M.~Strickland,
%``The Imaginary part of the static gluon propagator in an anisotropic
%(viscous) QCD plasma,''
Phys.\ Rev.\ D {\bf 79}, 114003 (2009).
%doi:10.1103/PhysRevD.79.114003
%[arXiv:0903.4703 [hep-ph]].
%%CITATION = doi:10.1103/PhysRevD.79.114003;%%

\bibitem{Burnier:2009yu}
  Y.~Burnier, M.~Laine and M.~Vepsalainen,
  %``Quarkonium dissociation in the presence of a small momentum space anisotropy,''
  Phys.\ Lett.\ B {\bf 678}, 86 (2009).
  %doi:10.1016/j.physletb.2009.05.067
  %[arXiv:0903.3467 [hep-ph]].
  %%CITATION = doi:10.1016/j.physletb.2009.05.067;%%
  %46 citations counted in INSPIRE as of 11 Nov 2016

\bibitem{Strickland:2011mw}
  M.~Strickland,
  %``Thermal $\upsilon_{1s}$ and chi_b1 suppression in $\sqrt{s_{NN}}=2.76$ TeV Pb-Pb collisions at the LHC,''
  Phys.\ Rev.\ Lett.\  {\bf 107}, 132301 (2011).
  %doi:10.1103/PhysRevLett.107.132301
  %[arXiv:1106.2571 [hep-ph]].
  %%CITATION = doi:10.1103/PhysRevLett.107.132301;%%
  %74 citations counted in INSPIRE as of 11 Nov 2016

\bibitem{Strickland:2011aa}
  M.~Strickland and D.~Bazow,
  %``Thermal Bottomonium Suppression at RHIC and LHC,''
  Nucl.\ Phys.\ A {\bf 879}, 25 (2012).
  %doi:10.1016/j.nuclphysa.2012.02.003
  %[arXiv:1112.2761 [nucl-th]].
  %%CITATION = doi:10.1016/j.nuclphysa.2012.02.003;%%
  %95 citations counted in INSPIRE as of 11 Nov 2016

\bibitem{Krouppa:2015yoa}
  B.~Krouppa, R.~Ryblewski and M.~Strickland,
  %``Bottomonia suppression in 2.76 TeV Pb-Pb collisions,''
  Phys.\ Rev.\ C {\bf 92}, no. 6, 061901 (2015).
  %doi:10.1103/PhysRevC.92.061901
  %[arXiv:1507.03951 [hep-ph]].

\bibitem{Krouppa:2016jcl}
  B.~Krouppa and M.~Strickland,
  %``Predictions for bottomonia suppression in 5.023 TeV Pb-Pb collisions,''
  Universe {\bf 2}, no. 3, 16 (2016).
  %doi:10.3390/universe2030016
  %[arXiv:1605.03561 [hep-ph]].

\bibitem{Ryu:2015vwa}
  S.~Ryu, J.-F.~Paquet, C.~Shen, G.~S.~Denicol, B.~Schenke, S.~Jeon and C.~Gale,
  %``Importance of the Bulk Viscosity of QCD in Ultrarelativistic Heavy-Ion Collisions,''
  Phys.\ Rev.\ Lett.\  {\bf 115}, no. 13, 132301 (2015).
  %doi:10.1103/PhysRevLett.115.132301
  %[arXiv:1502.01675 [nucl-th]].

\bibitem{Arnold:2006fz}
P.~B.~Arnold, C.~Dogan and G.~D.~Moore,
%``The Bulk Viscosity of High-Temperature QCD,''
Phys.\ Rev.\ D {\bf 74}, 085021 (2006).
%doi:10.1103/PhysRevD.74.085021
%[hep-ph/0608012].
%%CITATION = doi:10.1103/PhysRevD.74.085021;%%

\bibitem{Bazavov:2014pvz}
for recent results for $N_f=2+1$ QCD see for example
A.~Bazavov {\it et al.} [HotQCD Collaboration],
%``Equation of state in ( 2+1 )-flavor QCD,''
Phys.\ Rev.\ D {\bf 90}, 094503 (2014).
%doi:10.1103/PhysRevD.90.094503
%[arXiv:1407.6387 [hep-lat]].
%%CITATION = doi:10.1103/PhysRevD.90.094503;%%

\bibitem{KharzeevTuchin}
D.~Kharzeev and K.~Tuchin,
%``Bulk viscosity of QCD matter near the critical temperature,''
JHEP {\bf 0809}, 093 (2008);\\
%doi:10.1088/1126-6708/2008/09/093
%[arXiv:0705.4280 [hep-ph]].
%%CITATION = doi:10.1088/1126-6708/2008/09/093;%%
F.~Karsch, D.~Kharzeev and K.~Tuchin,
%``Universal properties of bulk viscosity near the QCD phase transition,''
Phys.\ Lett.\ B {\bf 663}, 217 (2008).
%doi:10.1016/j.physletb.2008.01.080
%[arXiv:0711.0914 [hep-ph]].
%%CITATION = doi:10.1016/j.physletb.2008.01.080;%%

\bibitem{qmass}
F.~R.~Brown, F.~P.~Butler, H.~Chen, N.~H.~Christ, Z.~h.~Dong,
W.~Schaffer, L.~I.~Unger and A.~Vaccarino,
%``On the existence of a phase transition for QCD with three light quarks,''
Phys.\ Rev.\ Lett.\  {\bf 65}, 2491 (1990);\\
%doi:10.1103/PhysRevLett.65.2491
%%CITATION = doi:10.1103/PhysRevLett.65.2491;%%
S.~Gavin, A.~Gocksch and R.~D.~Pisarski,
%``QCD and the chiral critical point,''
Phys.\ Rev.\ D {\bf 49}, R3079 (1994).
%doi:10.1103/PhysRevD.49.R3079
%[hep-ph/9311350].
%%CITATION = doi:10.1103/PhysRevD.49.R3079;%%

\bibitem{muB_critical_point}
J.~Berges and K.~Rajagopal,
%``Color superconductivity and chiral symmetry restoration at nonzero
%baryon density and temperature,''
Nucl.\ Phys.\ B {\bf 538}, 215 (1999);\\
%doi:10.1016/S0550-3213(98)00620-8
%[hep-ph/9804233].
%%CITATION = doi:10.1016/S0550-3213(98)00620-8;%%
A.~M.~Halasz, A.~D.~Jackson, R.~E.~Shrock, M.~A.~Stephanov and
J.~J.~M.~Verbaarschot,
%``On the phase diagram of QCD,''
Phys.\ Rev.\ D {\bf 58}, 096007 (1998);\\
%doi:10.1103/PhysRevD.58.096007
%[hep-ph/9804290].
%%CITATION = doi:10.1103/PhysRevD.58.096007;%%
M.~A.~Stephanov, K.~Rajagopal and E.~V.~Shuryak,
%``Signatures of the tricritical point in QCD,''
Phys.\ Rev.\ Lett.\  {\bf 81}, 4816 (1998).
%doi:10.1103/PhysRevLett.81.4816
%[hep-ph/9806219].
%%CITATION = doi:10.1103/PhysRevLett.81.4816;%%

\bibitem{Moore:2008ws}
G.~D.~Moore and O.~Saremi,
%``Bulk viscosity and spectral functions in QCD,''
JHEP {\bf 0809}, 015 (2008).
%doi:10.1088/1126-6708/2008/09/015
%[arXiv:0805.4201 [hep-ph]].
%%CITATION = doi:10.1088/1126-6708/2008/09/015;%%

\bibitem{Monnai:2016kud}
A.~Monnai, S.~Mukherjee and Y.~Yin,
%``Phenomenological Consequences of Enhanced Bulk Viscosity Near the
%QCD Critical Point,''
arXiv:1606.00771 [nucl-th].
%%CITATION = ARXIV:1606.00771;%%

\bibitem{Denicol:2014vaa}
G.~S.~Denicol, S.~Jeon and C.~Gale,
%``Transport Coefficients of Bulk Viscous Pressure in the 14-moment
%approximation,''
Phys.\ Rev.\ C {\bf 90}, no. 2, 024912 (2014).
%doi:10.1103/PhysRevC.90.024912
%[arXiv:1403.0962 [nucl-th]].
%%CITATION = doi:10.1103/PhysRevC.90.024912;%%

\bibitem{BulkShearKinetic}
G.~S.~Denicol, W.~Florkowski, R.~Ryblewski and M.~Strickland,
%``Shear-bulk coupling in nonconformal hydrodynamics,''
Phys.\ Rev.\ C {\bf 90}, no. 4, 044905 (2014);\\
%doi:10.1103/PhysRevC.90.044905
%[arXiv:1407.4767 [hep-ph]].
%%CITATION = doi:10.1103/PhysRevC.90.044905;%%
A.~Jaiswal, R.~Ryblewski and M.~Strickland,
%``Transport coefficients for bulk viscous evolution in the relaxation
%time approximation,''
Phys.\ Rev.\ C {\bf 90}, no. 4, 044908 (2014);\\
%doi:10.1103/PhysRevC.90.044908
%[arXiv:1407.7231 [hep-ph]].
%%CITATION = doi:10.1103/PhysRevC.90.044908;%%
D.~Bazow, U.~W.~Heinz and M.~Martinez,
%``Nonconformal viscous anisotropic hydrodynamics,''
Phys.\ Rev.\ C {\bf 91}, no. 6, 064903 (2015).
%doi:10.1103/PhysRevC.91.064903
%[arXiv:1503.07443 [nucl-th]].
%%CITATION = doi:10.1103/PhysRevC.91.064903;%%

\bibitem{Mrowczynski:2000ed}
S.~Mrowczynski and M.~H.~Thoma,
%``Hard loop approach to anisotropic systems,''
Phys.\ Rev.\ D {\bf 62}, 036011 (2000).
%doi:10.1103/PhysRevD.62.036011
%[hep-ph/0001164].
%%CITATION = doi:10.1103/PhysRevD.62.036011;%%

\bibitem{Carrington:1997sq}
M.~E.~Carrington, De-fu~Hou and M.~H.~Thoma,
%``Ward identities in nonequilibrium QED,''
Phys.\ Rev.\ D {\bf 58}, 085025 (1998);
%doi:10.1103/PhysRevD.58.085025
%[hep-th/9801103].
%%CITATION = doi:10.1103/PhysRevD.58.085025;%%
%M.~E.~Carrington, D.~f.~Hou and M.~H.~Thoma,
%``Equilibrium and nonequilibrium hard thermal loop resummation in the
%real time formalism,''
Eur.\ Phys.\ J.\ C {\bf 7}, 347 (1999).
%doi:10.1007/s100520050412, 10.1007/s100529800996
%[hep-ph/9708363].
%%CITATION = doi:10.1007/s100520050412, 10.1007/s100529800996;%%

\bibitem{Nopoush:2014pfa}
M.~Nopoush, R.~Ryblewski and M.~Strickland,
%``Bulk viscous evolution within anisotropic hydrodynamics,''
Phys.\ Rev.\ C {\bf 90}, no. 1, 014908 (2014).
%doi:10.1103/PhysRevC.90.014908
%[arXiv:1405.1355 [hep-ph]].
%%CITATION = doi:10.1103/PhysRevC.90.014908;%%

%\cite{Burnier:2014ssa}
\bibitem{Burnier:2014ssa}
  Y.~Burnier, O.~Kaczmarek and A.~Rothkopf,
  %``Static quark-antiquark potential in the quark-gluon plasma from lattice QCD,''
  Phys.\ Rev.\ Lett.\  {\bf 114}, no. 8, 082001 (2015)
  %doi:10.1103/PhysRevLett.114.082001
  %[arXiv:1410.2546 [hep-lat]].
  %%CITATION = doi:10.1103/PhysRevLett.114.082001;%%
  %36 citations counted in INSPIRE as of 25 Nov 2016

\bibitem{QQbar_HIC}
  A.~Mocsy, P.~Petreczky and M.~Strickland,
  %``Quarkonia in the Quark Gluon Plasma,''
  Int.\ J.\ Mod.\ Phys.\ A {\bf 28}, 1340012 (2013);\\
  %doi:10.1142/S0217751X13400125
  %[arXiv:1302.2180 [hep-ph]].
  %%CITATION = doi:10.1142/S0217751X13400125;%%
  A.~Andronic {\it et al.},
  %``Heavy-flavour and quarkonium production in the LHC era: from
  %proton–proton to heavy-ion collisions,''
  Eur.\ Phys.\ J.\ C {\bf 76}, no. 3, 107 (2016).
  %doi:10.1140/epjc/s10052-015-3819-5
  %[arXiv:1506.03981 [nucl-ex]].
  %%CITATION = doi:10.1140/epjc/s10052-015-3819-5;%%

\bibitem{Blaizot:1993zk}
J.~P.~Blaizot and E.~Iancu,
%``Kinetic equations for long wavelength excitations of the quark -
%gluon plasma,''
Phys.\ Rev.\ Lett.\  {\bf 70}, 3376 (1993);
%doi:10.1103/PhysRevLett.70.3376
%[hep-ph/9301236].
%%CITATION = doi:10.1103/PhysRevLett.70.3376;%%
%``Soft collective excitations in hot gauge theories,''
Nucl.\ Phys.\ B {\bf 417}, 608 (1994);\\
%doi:10.1016/0550-3213(94)90486-3
%[hep-ph/9306294].
%%CITATION = doi:10.1016/0550-3213(94)90486-3;%%
V.~P.~Nair,
%``Hamiltonian analysis of the effective action for hard thermal loops in QCD,''
Phys.\ Rev.\ D {\bf 50}, 4201 (1994)
%doi:10.1103/PhysRevD.50.4201
%[hep-th/9403146].
%%CITATION = doi:10.1103/PhysRevD.50.4201;%%


\end{thebibliography}
\end{document}